# Title: Observed Properties of Extrasolar Planets

Author: Andrew W. Howard[1]*

**Affiliations:**

[1]Institute for Astronomy, University of Hawai`i at Manoa, 2680 Woodlawn Drive, Honolulu, HI 96822, USA

*To whom correspondence should be addressed: howard@ifa.hawaii.edu

**Abstract**: Observational surveys for extrasolar planets probe the diverse outcomes of planet formation and evolution. These surveys measure the frequency of planets with different masses, sizes, orbital characteristics, and host star properties. Small planets between the sizes of Earth and Neptune substantially outnumber Jupiter-sized planets. The survey measurements support the core accretion model in which planets form by the accumulation of solids and then gas in protoplanetary disks. The diversity of exoplanetary characteristics demonstrates that most of the gross features of the Solar system are one outcome in a continuum of possibilities. The most common class of planetary system detectable today consists of one or more planets approximately one to three times Earth's size orbiting within a fraction of the Earth-Sun distance.

**Main Text:**

Extrasolar planets can be detected and characterized by high-resolution spectroscopy or precision photometry of the stars that they orbit. While some planetary systems have familiar properties, many have characteristics not seen in the solar system—small star-planet separations that result in planets heated to >1000 K, highly eccentric and inclined orbits, planets orbiting binary stars, and planet masses and sizes not represented locally. This diversity of planetary systems echoes the Copernican principle: the Earth is not the center of the universe and the solar system does not provide a universal template for planetary system architectures.

Measurements of the properties of a large number of planetary systems can probe the mechanisms of planet formation and place our solar system in context. These surveys can answer questions such as, what are common planet sizes and architectures and how did such planetary systems form? Measurements of extrasolar planets are mostly limited to gross physical properties including mass, size, orbital characteristics, and in some cases composition. Detailed measurements made in the solar system such as spatially-resolved imaging, in situ observations, and sample return are infeasible for extrasolar systems for the foreseeable future. Nevertheless, the sheer number of detected extrasolar planets compensates for the coarser measurements.

## Searching for Planets

This review focuses on planet populations that are detectable in large numbers by current transit and Doppler surveys: low-mass planets orbiting within about one astronomical unit (AU, the Earth-Sun distance) of their host stars and gas giant planets orbiting within several AU. The Doppler technique has detected and characterized ~700 planets orbiting ~400 stars (*1-2*). The *Kepler* space telescope has discovered more than 2700 planet candidates (*3-5*), of which only 5-10% are likely to be false positive detections (*6-7*). Giant planets in more distant orbits have also been detected by imaging (*8*) and gravitational microlensing surveys (*9*).

With the Doppler technique, planet masses and orbits are inferred from the observed motions of their host stars. Stellar orbits are point reflections of their planets' orbits, scaled down by the planet-to-star mass ratios. These orbits are measured by the star's line-of-sight velocity (radial velocity, RV) using high-resolution spectroscopy with ground-based telescopes. Planets are detected by analyzing the repeating patterns in the time series RV measurements and are characterized by their orbital periods ($P$), minimum masses ($M\sin i$, where $M$ is a planet mass and $i$ is the inclination of a planet's orbit relative to the sky plane), and orbital eccentricities (Fig. 1). Planets with larger masses and shorter orbital periods orbiting lower mass stars are more detectable. Sensitivity to planets varies from star to star and depends on details of the observing history, including the number, precision, and time baseline of the RV measurements. The earliest Doppler surveys began 20-25 years ago with a few hundred nearby, bright stars and are now sensitive to analogs of Jupiter and Saturn. With measurement precisions of ~1 m s$^{-1}$, more recent surveys are detecting planets of a few Earth masses ($M_E$) for close-in orbits.

With the transit technique, the eclipses of planets whose orbits happen to be viewed edge-on are detected as brief dips in the host star's brightness (Fig. 1). The size of the planet relative to the star is inferred from the depth of the transit. Jupiter-size planets block ~1% of the flux from Sun-like stars and are detectable using ground-based telescopes, whereas the 0.01%-deep transits of Earth-size planets are only detectable by precise, space-borne telescopes such as *Kepler*. The planet's orbital period is the time interval between consecutive transits and the orbital distance (semi-major axis) can be inferred from Kepler's third law. The mass of a transiting planet can be measured from follow-up Doppler observations if the host star is bright enough and the Doppler amplitude large enough. Masses can also be measured in special cases from precise timing of consecutive transits, which deviate from strict periodicity when multiple planets orbiting the same star gravitationally perturb one another (10-11).

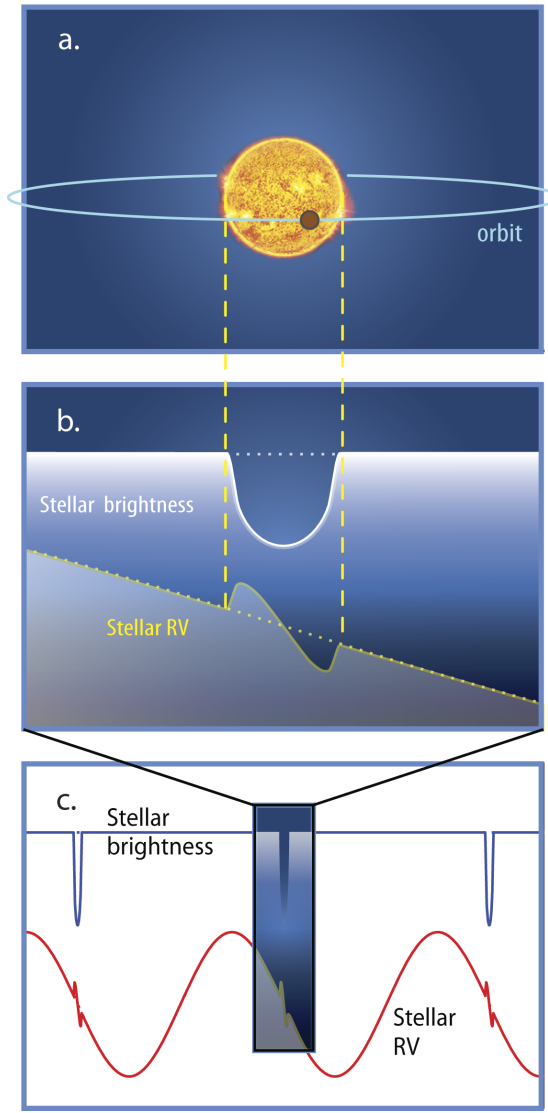

**Fig. 1**. Schematic illustration of a planetary orbit and the variations in stellar brightness and radial velocity (RV) that it causes. **(A)** A planet orbits its host star and eclipses ('transits') the star as seen by a distant observer. **(B)** Planets are detectable during transit by the decrease in stellar brightness (solid white line). Transit depth is proportional to the blocked fraction of the stellar disk. The stellar obliquity can be measured during transit by anomalous Doppler shifts (the Rossiter-McLaughlin effect; solid yellow line) in the RV time series as the planet blocks portions of the rotating stellar disk. A low-obliquity system with a well-aligned stellar spin axis and planet orbital axis is shown. Non-transiting planets do not produce such effects (dashed lines). **(C)** Over many orbits, planet properties including the size, mass, orbital period, eccentricity, and orbital inclination can be measured from detailed analysis of time-series photometric and RV data.

Transiting planets also offer the opportunity to measure the stellar obliquity—the angle between the stellar spin axis and a planet's orbital axis (*12*)—as well as characteristics of the planet's atmosphere (*13*). Obliquities have primarily been measured as the transiting planet alternately blocks blue-shifted and red-shifted portions of the rotating stellar disk causing apparent Doppler shifts (the Rossiter-McLaughlin effect, see Fig. 1). Obliquity measurements are sensitive to past dynamical interactions that can perturb planets into highly-inclined orbits.

Taken together, Doppler and transit detections probe the bulk physical properties of planets (masses, radii, and densities) and their orbital architectures (the number of planets per system and their orbital separations, eccentricities, and geometries). In an observational survey, a large number of stars are searched for planets and the statistical properties of the detected population are studied to infer mechanisms of planet formation and evolution. Surveys count planets and naturally produce number distributions of planet parameters (e.g. the number of detected planets versus planet mass), but these distributions can systematically hide planets that are more difficult to detect. To measure planet occurrence—how commonly planets with a particular characteristic exist in nature—surveys must estimate their sensitivity to planets with different values of that characteristic and statistically correct for their incomplete sample of detected planets.

**Abundant close-in, low-mass planets**

Planets intermediate in size between Earth and Neptune are surprisingly common in extrasolar systems, but notably absent from our solar system. The planet size and mass distributions (Fig. 2) show clearly that these small planets outnumber large ones, at least for close-in orbits. Two separate Doppler surveys (14-15) of nearby, Sun-like stars have shown that planet occurrence rises significantly with decreasing planet mass ($M\sin i$) from 1000 $M_E$ (3 Jupiter masses) down to 3 $M_E$. In the Eta-Earth Survey, an unbiased set of 166 nearby, G and K-type stars in the Northern sky were observed at Keck Observatory (*14*). The RV of each star was measured dozens of times over five years and the time series RVs were searched for the signatures of planets with orbital periods $P < 50$ days. (The restriction of $P < 50$ days for solar-type stars is equivalent to restricting orbital distances to $< 0.25$ AU, inside of Mercury's 0.4 AU orbit.) In total, 33 planets were detected orbiting 22 of the 166 stars. Low-mass planets ($M\sin i = 3\text{-}30\ M_E$) were detected more frequently, in spite of their weaker Doppler signals. After correcting for survey incompleteness, the planet mass distribution was fit with a power law model that rises steeply toward low mass. The probability of a star hosting a close-in planet scales as $(M\sin i)^{-0.48}$. In absolute terms, 15% of Sun-like stars host one or more planets with $M\sin i = 3\text{-}30\ M_E$ orbiting within 0.25 AU, and by extrapolation another 14% of stars host planets with $M\sin i = 1\text{-}3\ M_E$.

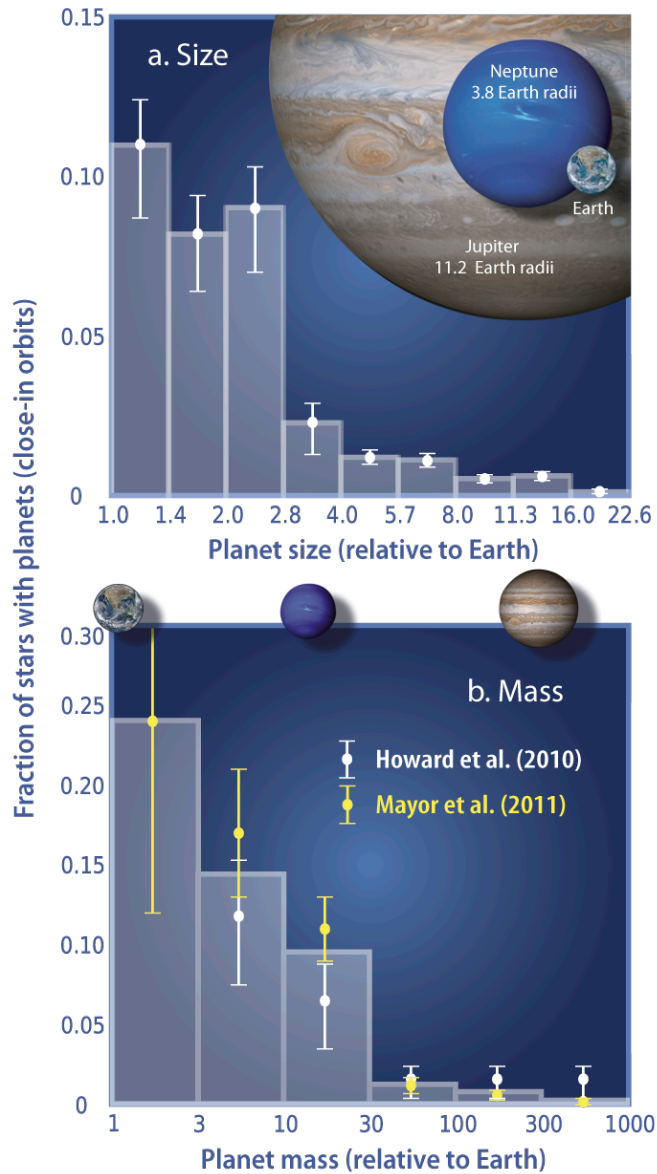

**Fig. 2**. The size **(A)** and mass **(B)** distributions of planets orbiting close to G and K-type stars. The distributions rise substantially with decreasing size and mass, indicating that small planets are more common than large ones. Planets smaller than 2.8 $R_E$ or less massive than 30 $M_E$ are found within 0.25 AU of 30-50% of Sun-like stars. **(A)** The size distribution from transiting planets shows occurrence versus planet radius and is drawn from two studies of *Kepler* data: (*16*) for planets smaller than four times Earth size and (*17-18*) for larger planets. The inset images of Jupiter, Neptune, and Earth show their relative sizes. The mass ($M\sin i$) distributions **(B)** show the fraction of stars having at least one planet with an orbital period shorter than 50 days (orbiting inside of ~0.25 AU) are from the Doppler surveys from (*14*, white points) and (***Error! Reference source not found.***, yellow points), whereas the histogram shows their average values. Inset images of Earth, Neptune, and Jupiter are shown on the top horizontal axis at their respective masses. Both distributions are corrected for survey incompleteness for small/low-mass planets to show the true occurrence of planets in nature.

The HARPS (High Accuracy Radial velocity Planet Searcher) survey measured the RVs of 376 Sun-like stars in the Southern sky with somewhat better sensitivity to low-mass planets (15). It confirmed the rising planet mass function with decreasing mass and extended it to 1-3 $M_E$ planets (Fig 2). It also demonstrated that low-mass planets have small orbital eccentricities and are commonly found in multi-planet systems with 2-4 small planets orbiting the same star with orbital periods of weeks or months. The HARPS survey found that at least 50% of stars have one or more planet of any mass with $P < 100$ days.

The *Kepler* mission has significantly refined our statistical knowledge of planets between the size of Earth and Neptune by revealing thousands of these planets, compared to the dozens detected by the Doppler technique. The distribution of planet sizes (radii) measured by *Kepler* (Fig. 2) follows the same trend as the mass distribution, with small planets being more common (16, 17, 19). However, the *Kepler* size distribution extends confidently down to Earth size for close-in planets, whereas the mass distribution is uncertain at the 50% level near one Earth mass. The size distribution is characterized by a power-law rise in occurrence with decreasing size (17) down to a critical size of ~2.8 Earth radii ($R_E$), below which planet occurrence plateaus (16). Earth size planets orbiting within 0.25 AU of their host stars are just as common as planets twice that size. The small planets detected by *Kepler* (<2 $R_E$) appear to have more circular orbits than larger planets (20), suggesting reduced dynamical interactions.

*Kepler* is sensitive to sub-Earth-size planets around stars with low photometric noise and has detected planets as small as the Moon (0.3 $R_E$) (21). However, survey sensitivity remains uncertain below 1 $R_E$ with current data, even for orbital periods of a few days. While the occurrence plateau for 1-2.8 $R_E$ with a steep fall-off for larger planets is not well understood theoretically, it offers an important observed property of planets around Sun-like stars that must be reproduced by planet formation models.

The high occurrence of small planets with $P < 50$ days likely extends to more distant orbits. As *Kepler* accumulates photometric data, it becomes sensitive to planets with smaller sizes and longer orbital periods. Based on 1.5 years of data, the small planet occurrence distribution as a function of orbital period is flat out to $P = 250$ days (with higher uncertainty for larger $P$). That is, the mean number of planets per star per logarithmic period interval is proportional to $P^{+0.11\pm0.05}$ and $P^{-0.10\pm0.12}$ for 1-2 $R_E$ and 2-4 $R_E$ planets, respectively (22).

Of the *Kepler* planet host stars, 23% show evidence for two or more transiting planets (3). To be detected, planets in multi-transiting systems likely orbit in nearly the same plane, with mutual inclinations of a few degrees at most, as in the solar system. The true number of planets per star (transiting or not) and their mutual inclinations can be estimated from simulated observations of model planetary systems constrained by the number of single, double, triple, etc. transiting systems detected by *Kepler* (23, 24). One model finds an intrinsic multi-planet distribution with 54%, 27%, 13%, 5%, and 2% of systems having 1, 2, 3, 4, and 5 planets with $P < 200$ days. Most multi-planet systems (85%) have mutual inclinations of less than 3° (23). Comparisons of the *Kepler* and HARPS planetary systems also suggest mutual inclinations of a few degrees (25). This high degree of co-planarity is consistent with planets forming in a disk without substantial dynamical perturbations capable of increasing inclinations.

Orbital period ratios in multi-transiting systems provide additional dynamical information. These ratios are largely random (26), with a modest excess just outside of period ratios that are consistent with dynamical resonances (ratios of 2:1, 3:2, etc.) and a compensating deficit inside

(e.g. *27*). The period ratios of adjacent planet pairs demonstrate that >31, >35, and >45% of 2-planet, 3-planet, and 4-planet systems are dynamically packed; adding a hypothetical planet would gravitationally perturb the system into instability (*28*).

**Masses, radii, and densities**

While mass and size distributions provide valuable information about the relative occurrence of planets of different types, it remains challenging to connect the two. Knowing the mass of a planet does not specify its size, and vice versa. A planet the mass of Earth could have a variety of sizes, depending on the composition and the extent of the atmosphere.

This degeneracy can be lifted for ~200 planets with well-measured masses and radii (Fig. 3), most of which are giant planets. The cloud of measurements in Fig. 3 follows a diagonal band from low-mass/small-size to high-mass/large-size. This band of allowable planet mass/size combinations has considerable breadth. Planets less massive than ~30 $M_E$ vary in size by a factor of 4-5 and planets larger than ~100 $M_E$ (gas giants) vary in size by a factor of ~2. For the gas giants, the size dispersion at a given mass is due largely to two effects. First, the presence of a massive solid core (or distributed heavy elements) increases a planet's surface gravity, causing it be more compact. Second, planets in tight orbits receive higher stellar flux and are statistically more likely to be inflated relative to the sizes predicted by atmospheric models (the "hydrogen" curve in Fig. 3). While it is clear that higher stellar flux correlates with giant planet inflation (*29*), it is unclear how the stellar energy is deposited in the planet's interior. Energy deposited in a planet's outer layers is quickly re-radiated unless it is somehow circulated to the interior.

Low-mass planets have an even greater variation in size and inferred compositional diversity. The planet Kepler-10b has a mass of 4.6 $M_E$ and a density of 9 g cm$^{-3}$, indicating a rock/iron composition. With such a high density, this planet likely has little or no atmosphere (*30*). In contrast, the planet Kepler-11e has a density of 0.5 g cm$^{-3}$ and a mass of 8 $M_E$. A substantial light-element atmosphere (probably hydrogen) is required to explain its mass and radius combination (*31*). The masses and radii of intermediate planets lead to ambiguous conclusions about composition. For example, the bulk physical properties of GJ 1214b (mass 6.5 $M_E$, radius 2.7 $R_E$, density 1.9 g cm$^{-3}$) are consistent with a several compositions: a "super-Earth" with a rock/iron core surrounded by ~3% $H_2$ gas by mass; a "water world" planet consisting of a rock/iron core, a water ocean and atmosphere that contribute ~50% of the mass; or a "mini-Neptune" composed of rock/iron, water, and H/He gas (*32*). For this particular planet, measuring the transmission spectrum during transit appears to have lifted the degeneracy. The small atmospheric scale height of GJ1214b favors a high mean molecular weight atmosphere (possibly water), but is also consistent with an $H_2$ or H/He atmosphere rendered featureless by thick clouds (*33*).

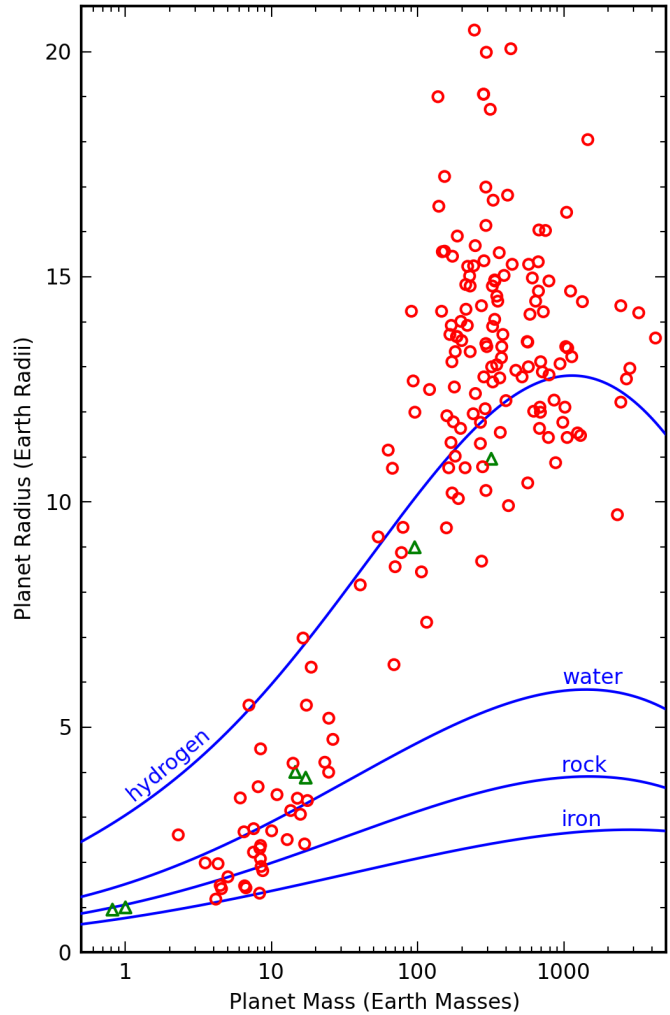

**Fig. 3**. Masses and sizes of well-characterized planets. Extrasolar planets (*1*, *35*, *36*) are shown as open red circles whereas solar system planets are designated by open green triangles. Radii are measured by transit photometry and masses are measured by radial velocity or transit timing methods. Model mass-radius relationships for idealized planets consisting of pure hydrogen (*37*), water, rock ($Mg_2SiO_4$), or iron (*38*) are shown as blue lines. Poorly understood heating mechanisms inflate some gas giant planets (larger than ~8 $R_E$) to sizes larger than predicted by the simple hydrogen model. Smaller planets (less massive than ~30 $M_E$) show great diversity in size at a fixed mass, likely due to varying density of solids and atmospheric extent. Gas giant planets are overrepresented relative to their occurrence in nature due to their relative ease of detection and characterization.

**Gas giant planets**

The orbits of giant planets are the easiest to detect using the Doppler technique and were the first to be studied statistically (e.g. *39*, *40*). Observations over a decade of a volume-limited sample of ~1000 F, G, and K-type dwarf stars at Keck Observatory showed that 10.5% of G and K-type dwarf stars host one or more giant planets (0.3-10 Jupiter masses) with orbital periods of 2-2000 days (orbital distances of ~0.03-3 AU). Within those parameter ranges, less massive and more distant giant planets are more common. A fit to the giant planet distribution in the mass-period plane shows that occurrence varies as $M^{-0.31\pm0.2}P^{+0.26\pm0.1}$ per logarithmic interval $d\log M\ d\log P$. Extrapolation of this model suggests that 17-20% of such stars have giant planets orbiting within 20 AU ($P = 90$ years) (*41*). This extrapolation is consistent with a measurement of giant planet occurrence beyond ~2 AU from microlensing surveys (*42*). However, the relatively few planet detections from direct imaging planet searches suggest that the extrapolation is not valid beyond ~65 AU (*43*).

These overall trends in giant planet occurrence mask local pile-ups in the distribution of orbital parameters (*44*). For example, the number distribution of orbital distances for giant planets shows a preference for orbits larger than ~1 AU and to a lesser extent near 0.05 AU, where "hot Jupiters" orbit only a few stellar radii from their host stars (Fig. 4a). This "period valley" for apparently single planets is interpreted as a transition region between two categories of planets with different migration histories (*39*). The excess of planets starting at ~1 AU approximately coincides with the location of the ice line. Water is condensed for orbits outside of the ice line, providing an additional reservoir of solids that may speed the formation of planet cores or act as a migration trap for planets formed farther out (*45*). The orbital period distribution for giant planets in multi-planet systems is more uniform, with hot Jupiters nearly absent and a suppressed peak of planets in >1 AU orbits. The giant planet eccentricity distribution (Fig. 4b) also differs between single and multi-planet systems. The eccentricity distribution for single planets can be reproduced quantitatively by a dynamical model in which initially low eccentricities are excited by planet-planet scattering (*46*). Multi-planet systems likely experienced substantially fewer scattering events. One interpretation is that eccentric, single planet systems are the survivors of scattering events that ejected the other planets in the system.

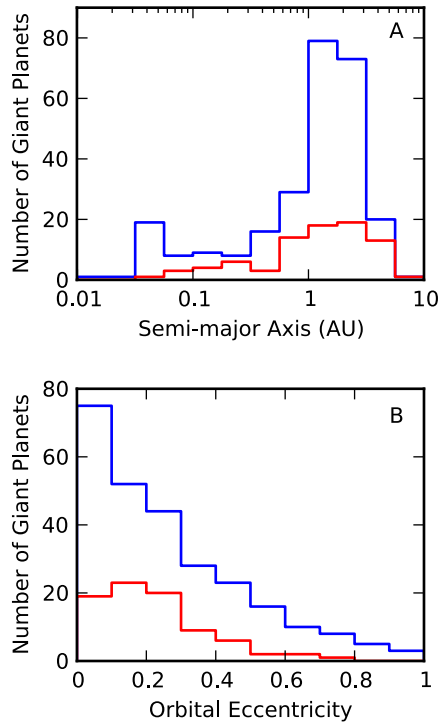

**Fig. 4**. Orbital characteristics of giant planets ($M\sin i > 0.2$ Jupiter masses) detected by Doppler surveys (*1,36*). The number distribution of semi-major axes **(A)** shows that apparently single planets (blue) preferentially orbit at distances of ~0.05 AU and at ~1-3 AU from their host stars. These preferred orbits are diminished in multi-planet systems (red). The decline in number of detected planets for orbits outside of ~3 AU is not significant; fewer stars have been searched for such planets compared to the closer orbits. The distribution of orbital eccentricities **(B)** for apparently single planets (blue) span the full range, with low-eccentricity orbits being more common. Giant planets in multi-planet systems (red) have orbits that are more commonly close to circular (eccentricity = 0). The larger eccentricities of single planets suggests that they were dynamically excited from a quiescent, nearly circular origin, perhaps by planet-planet scattering that resulted in the ejection of all but one detectable planet per system.

Although hot Jupiters (giant planets with $P < {\sim}10$ days) are found around only 0.5-1.0% of Sun-like stars (*47*), they are the most well-characterized planets because they are easy to detect and follow up with ground- and space-based telescopes. However, they have a number of unusual characteristics and a puzzling origin. In contrast to close-in small planets, hot Jupiters are commonly the only detected planet orbiting the host star within observational limits (*48*; see also Fig. 4). Many have low eccentricities, primarily due tidal circularization. The measured obliquities of stars hosting hot Jupiters display a peculiar pattern: obliquities are apparently random above a critical stellar temperature of ~6250 K, but cooler systems are mostly aligned. In situ formation is unlikely for hot Jupiters because of insufficient protoplanetary disk mass so close to the star. Rather, they likely formed in the disk at several AU, were gravitationally perturbed into orbits with random inclinations and high eccentricities, and were captured at ~0.05 AU by dissipation of orbital energy in tides raised on the planet. For systems with sufficiently strong tides raised by the planet on the star (which depend on a stellar convective zone that is only present below ~6250 K), the stellar spin axis aligns to the orbital axis (*12*).

**Planet formation**

Metal-rich stars (49) are more likely to host giant planets within 5 AU. This "planet-metallicity correlation" was suggested in 1997 when the first four Doppler-discovered giant planets were all found to orbit stars with higher iron abundance than the Sun (*50*). Initially, this correlation was seen as an artifact of stellar self-pollution from accretion of metals onto the stellar atmosphere during planet formation. Today, it provides evidence for the core accretion model of planet formation (51). In this model, a high density of solids in the protoplanetary disk is required to form giant planets, which pass through two key phases. Protoplanets must grow to masses of ~5-10 $M_E$ by accretion of solids (dust and ice) from the disk. The protoplanet then undergoes runaway gas accretion, increasing its mass by an order of magnitude, but only if the protoplanetary gas has not yet dissipated. Giant planet formation is a race against dispersal of gas from the protoplanetary disk on a timescale of ~3-5 Myr (see (*52*) for a review). This theory predicts that giant planets should be more common around massive and metal-rich stars whose disks have higher surface densities of solids.

The planet-metallicity correlation was validated statistically by Doppler surveys of ~1000 stars with masses of 0.7-1.2 solar masses ($M_{Sun}$) and uniformly-measured metallicities, which were highly sensitive to giant planets (*53*, *54*). The probability of a star hosting a giant planet is proportional to the square of the number of iron atoms in the star relative to the Sun, $p(\text{planet}) \propto N_{Fe}^2$ (53). A later Doppler study spanned a wider range of stellar masses (0.3-2.0 $M_{Sun}$) and showed that the probability of a star hosting a giant planet correlates with both stellar metal content and stellar mass, $p(\text{planet}) \propto N_{Fe}^{1.2\pm0.2} M_{star}^{1.0\pm0.3}$ (*55*). The planet-metallicity correlation only applies to gas giant planets. Planets larger than 4 $R_E$ (Neptune size) preferentially orbit metal-rich stars, whereas smaller planets are found in equal numbers around stars with a broad range of metallicities (*56*). That is, small planets form commonly in most protoplanetary disks, but only a fraction grow to a critical size in time to become gas giants.

Although the planet-metallicity trends support the basic mechanism of the core accretion model, many statistical features of the observed planet population cannot yet be explained in detail. In particular, the population of low-mass planets inside of 1 AU is difficult to reproduce in conventional models. Population synthesis models attempt to follow the growth and migration of sub-Earth-size protoplanets in a protoplanetary disk to predict the planet masses and orbital distances after the disk dissipates (e.g. *45, 57*). These models reproduce the giant planet population well, but struggle with low-mass planets. In population synthesis models, low-mass planets form primarily near and beyond the ice line (several AU) and migrate to close orbital distances by interactions with the disk. The prescriptions for migration and growth in these models produce "deserts" of reduced planet occurrence precisely where Doppler and transit surveys detect a great abundance of planets.

An alternative model is in situ formation of close-in, low-mass planets with minimal subsequent planet migration (e.g. *58, 59*). While this model correctly reproduces several observed properties of close-in planets (the mass distribution, multi-planet frequencies, small eccentricities and inclinations), it is still in the early stages of development. One challenge is that in situ formation requires ~20-50 $M_E$ of protoplanetary disk material inside of 1 AU, which is poorly constrained by observations.

**Earth-size planets in the Habitable Zone**

The detection of planets the size or mass of Earth remains a prominent observational goal. Using the Doppler technique, one such planet has been detected: a planet with $M\sin i$ = 1.1 $M_E$ orbiting the nearby star α Centauri B with an orbital separation of 0.04 AU that renders it too hot for life (*60*). The Doppler signal from an Earth-mass planet orbiting at 1 AU is a factor of five smaller, beyond the reach of current instrumentation and possibly hidden in Doppler noise from the star. Nevertheless, Doppler planet searches continue. If a planet the size or mass of the Earth is detected orbiting in the habitable zone of a nearby star, it would be a milestone for science and could catalyze the development of instruments to image and take spectra of such planets.

*Kepler* has detected dozens of Earth-size planets, although these planets orbit interior to their stars' habitable zones (16, *19*). The habitable zone—the set of planetary orbits consistent with liquid water existing on the planet's surface—is challenging to define precisely because it depends on the detailed energy balance for a planet with an often poorly constrained composition and atmosphere (*61*). Nevertheless, *Kepler* has also detected planets slightly larger than Earth (1.4 and 1.6 times Earth size) in the classically defined habitable zone (*62*). The primary goal for *Kepler* in its extended mission is to detect individual Earth-size planets in the habitable zone and to estimate their occurrence rate. Not all of the Earth-size planets detected by *Kepler* will be one Earth mass, however. Measuring the densities of several Earth-size planets (not necessarily in the habitable zone) will offer some constraints on the typical compositions of Earth-size planets.

Targeting low-mass stars offers a shortcut in the search for planets the size and mass of Earth. Planets are more detectable in Doppler and transit searches of lower mass stars. The habitable zones around such stars are also closer, owing to the reduced brightness of low-mass stars. Small planets may be more common around low-mass stars as well (*17*, but see also *19*). An analysis of the *Kepler* planets orbiting M dwarfs suggests a high rate of overall planet occurrence, 0.9 planets per star in the size range 0.5-4 $R_E$ in $P$ < 50 day orbits. Earth-size planets (0.5-1.4 $R_E$) are found in the habitable zones of $15^{+13}_{-6}$% of the M dwarfs in the *Kepler* sample (*63*). As *Kepler*'s sensitivity expands during the extended mission, we will likely learn how common Earth-size planets are in the habitable zones of Sun-like stars.

**References and Notes:**

1. J. T. Wright *et al., Proc. Astron. Soc. Pacific*, **123**, 412 (2011)
2. Updates to (*1*) are available at http://exoplanets.org

**Acknowledgments:** I thank K. Teramura for assistance preparing the graphics and G. Marcy, D. Fischer, J. Wright, T. Currie, E. Petigura, H. Isaacson, R. Dawson, I. Crossfield, S. Kane, J. Steffen, J. Maurer, and S. Howard for comments on this manuscript. I also thank three anonymous reviewers for constructive criticism. I acknowledge partial funding from NASA grant NNX12AJ23G.